\documentclass[aps,preprint,showpacs]{revtex4}
\usepackage{psfig}
\begin{document}        %
\draft
\title{Theoretical Study for Deformation Kinetics of \\ Glassy Solid Helium   within Cylindrical Microtubes}  %
\author{Zotin K.-H. Chu} 
\affiliation{3/F, 4, Alley 2, Road Xiushan, Leshanxinchun,
Xujiahui 200030}
%
\begin{abstract}
We study the  deformation kinetics for the (glassy) solid helium
confined in microscopic domain at very low temperature regime by
using an absolute-reaction-rate model considering the shear
thinning behavior which means, once material being subjected to
high shear rates, the viscosity diminishes with increasing shear
rate. Our calculations show that there might be nearly
frictionless fields for rate of deformation due to the almost
vanishing shear stress in microtubes at very low temperature
regime together with wavy-rough corrugations along micropores and
the slip. As the tube size decreases, the surface-to-volume ratio
increases and therefore, surface roughness will greatly affect the
deformation kinetics in micropores. After using the boundary
perturbation method, we have obtained a class of temperature and
activation energy dependent fields for the deformation kinetics at
low temperature regime with the presumed small wavy roughness
distributed along the wall of an cylindrical microtube. The
critical  deformation kinetics of the glassy matter is dependent
upon the temperature, activation energy, activation volume,
orientation dependent and is proportional to the (referenced)
shear rate, the slip length, the amplitude and the orientation of
the wavy-roughness. Finally, we also discuss the quantitative
similarity between our results with Ray and Hallock [Phys. Rev.
Lett. {\bf 100}, 235301 (2008)].
%
%
%
\end{abstract}
%
\maketitle
\bibliographystyle{plain}
\section{Introduction}
Grigor'ev {\it et al.} recently  performed high-precision pressure
measurement in solid $^4$He samples grown by capillary blocking
technique. In all their nonannealed hcp crystals, the temperature
dependence of pressure demonstrates a contribution proportional to
T$^2$, the latter becomes the leading term at temperatures T
$<300$ mK, at which {\it supersolid} effects were observed.
Grigor'ev {\it et al.} thus claimed that such a behavior may be
ascribed to a glassy phase. They also found that this glassy
contribution to pressure can be eliminated only by a substantial
annealing : A dramatic pressure decrease of $\sim 2$ bar was
observed under annealing at temperatures very close to the melting
point. They thus conjectured that this effect is due to
solidification of liquid or glass captured in closed cavities
during the growth process [1].
\newline
Meanwhile Day and Beamish recently observed an approximately 10\%
increase in the shear modulus of $^4$He at low temperatures (below
200 mK) [2].  Ray and  Hallock found evidence for flow through
solid $^4$He in at least some cases [3]. To be specific, Ray and
Hallock have conducted experiments that show the first evidence
for flow of helium through a region containing solid hcp $^4$He
off the melting curve. Their phase diagram appears to have two
regions. Samples grown at lower pressures show flow, with flow
apparently dependent on sample history, with reduced flow for
samples at higher temperature, which is evidence for dependence on
temperature. Samples grown at higher pressures show no clear
evidence for any such flow for times longer than $10$ hours [3].
The temperatures utilized for their work are well above the
temperatures at which much attention has been focused, but
interesting behavior was seen.
\newline
Note also that Clark {\it et al.}  have studied the thermal
history of the resonant frequency of a torsional oscillator
containing solid $^4$He. They found that the magnitude of the
frequency shift that occurs below $\sim 100$ mK is multi-valued in
the low-temperature limit, depending strongly on how the state is
prepared [4]. However, Anderson interprets the observed NCRI
(nonclassical rotational inertia, please see [4] for the detailed
references) as a consequence of vortex liquid and supercurrents
flowing around a thermally excited fluctuation of vortices [5].
The subsequent experiments [6] reported the first observations on
the time-dependent dissipation when the drive level of a torsional
oscillator containing solid $^4$He was abruptly changed. The
relaxation of dissipation in solid $^4$He shows rich dynamical
behavior including exponential and logarithmic time-dependent
decays, hysteresis, and memory effects. As remarked in [6], their
procedure of initiating the oscillation at a high temperature is
likely to have brought the sample solid $^4$He to a
quasi-equilibrium array of vortices for the drive level at the
temperature of measurement. The observed logarithmic time
dependence probably arises from the {\it thermally activated
motion} of vortices from one metastable state to another. This
process continues throughout the measurement while the sample was
maintained at the constant temperature. Just after the drive level
has been decreased, the existing vortices are out of equilibrium
with the new drive level. The observed undershoot may be a
transient excess dissipation owing to the 'wrong' number of
vortices. The vortices then must adjust to the new drive level by,
say, moving out of the sample. The characteristic time involved in
the macroscopic motion of the vortices was then the observed time
constant [6]. The motion presumably involves processes occurring
both within the sample and at the surface boundaries. The time
constant would become shorter as the vortices begin to move more
freely. The proposed vortex liquid state [5] appears to occur
above about $60$ mK. When the vortex liquid 'freezes', the time
constant would diverge [6].
\newline
Researches mentioned above all imply that the supersolid
(material) is not so simple but really complex. We also noticed
that as reported by Dyre {\it et al.} in 1996 for  glass-forming
molecular liquids the (high-frequency) shear modulus increases as
the temperature decreases [7]. The latter resembles that reported
in [2] by Day and Beamish. Thus, it's necessary to study the
deformation kinetics as well as transport of amorphous and/or
glassy material (presumed solid helium to be almost the same)
under confined microdomains at rather low temperature regime!
\newline
 Glasses are amorphous materials of polymeric, metallic,
inorganic or organic type. The plastic deformation of amorphous
materials and glasses at low temperatures and high strain rates is
known to be inhomogeneous and rate-dependent. In fact, the
mechanical behavior of amorphous materials such as  bulk metallic
glasses [7-10] continues to present great theoretical challenges.
While dislocations have long been recognized as playing a central
role in plasticity of crystalline systems, no counterpart is
easily identifiable in disordered matter. In addition, yield and
deformation kinetics [11-12] occur very far from equilibrium,
where the state of the system may have a complex history
dependence.
\newline In recent years, considerable effort was geared towards
understanding how glasses respond to shear [12]. Phenomena such as
shear thinning and 'rejuvenation' are common when shear
deformation (rate) is imposed. At low temperatures they behave in
a brittle elastic manner; at high temperatures, much above the
glass transition the behavior is more (rubbery) (viscoelastic).
There is a huge drop in modulus when the temperature is increased
above the glass transition temperature, indicating a shift in
behavior from (glassy) to (rubbery). Because of these peculiar
mechanical properties of amorphous materials, the linear theory of
viscoelasticity is unable to model closely the observed response
and thus there is a need for a non-linear theory of
viscoelasticity. \newline Unlike crystals, glasses also age,
meaning that their state depends on their history. When a glass
falls out of equilibrium, it evolves over very long time scales.
Motivated by the above issues and the interesting characteristics
of deformation kinetics at very low temperature  we shall study
the deformation kinetics in microscopic domain at low enough
temperature which is an interesting topic for applications in
micro- and nanodomains  or the validation in using quantum
mechanic formulations [13] where the nonlinear constitutive
relations should be adopted.
\newline
However, real surfaces are rough  at the micro- or
even at the meso-scale 
and the role of surface roughness has been extensively
investigated, and opposite conclusions have been reached so far
[14-15]. For instance, friction can increase when two opposing
surfaces are made smoother (this is the case of cold welding of
highly polished metals). On the other hand, friction increases
with roughness when interlocking effects among the asperities come
into play. This apparent contradiction is due to the effects of
length scales, which appear to be of crucial importance in this
phenomenon.
\newline From the mechanical point of view, a contact problem
involves the determination of the traction distributions
transmitted from one surface to the other, in general involving
normal pressures and, if friction is present, shear tractions,
according to an appropriate set of equalities and inequalities
governing the physics of the contact [16]. When there is friction
at the contact interface, Coulomb friction behaviour is usually
introduced to give the conditions necessary to determine the shear
traction distribution. Any point in the contact area must be
either in 'stick', or 'slip' condition, and the tangential
tractions must behave accordingly.\newline
In this paper we shall consider the deformation kinetics of
amorphous solid helium at very low temperature in micropores which
have radius- or transverse-corrugations along the cross-section.
The glassy matter  will be treated as a shear-shinning material.
To consider the transport of this kind of glass (shear-thinning)
material in microdomains, we adopt the verified model initiated by
Cagle and Eyring [9] which was used to study the annealing of
glass. To obtain the law of annealing of glass for explaining the
too rapid annealing at the earliest time, because the relaxation
at the beginning was steeper than could be explained by the
bimolecular law, Cagle and Eyring [9] tried a hyperbolic sine law
between the shear (strain) rate : $\dot{\Gamma}$ and (large) shear
stress : $\tau$ and obtained the close agreement with experimental
data. This model has sound physical foundation from the thermal
activation process (Eyring [10] already considered a kind of
(quantum) tunneling which relates to the matter rearranging by
surmounting a potential energy barrier; cf., Fig. 1) and thus it
might also resolve the concern raised by Anderson [5] for the
thermal noises to the superflow of vortex liquid (i.e., the
supersolid helium). \newline With this model we can associate the
(glassy) matter with the momentum transfer between neighboring
atomic clusters on the microscopic scale and reveals the atomic
interaction in the relaxation of flow with (viscous) dissipation.
\newline The outline of this short paper is as follows. Section 2
describes the general  physical formulations of the framework. In
this Section, explicit derivations  for the glassy deformation
kinetics are introduced based on a  microscopic model proposed by
Eyring [10]. The boundary perturbation technique [17] will be
implemented, too. In the third Section, we consider the very-low
temperature limit of  our derived solutions which are highly
temperature as well as activation energy dependent at rather low
temperature regime. Relevant results and discussion are given
therein.
\section{Theoretical formulations}
The beginnings of theoretical molecular mechanisms of deformation
in amorphous materials and glass are as old as the subject of
atomic mechanisms of deformation and yield in metals. The first
specific molecular mechanism of deformation for amorphous
materials and glass was published by Eyring [10] and later, Taylor
[18] published the model of an edge dislocation to account for the
plastic deformation in metals. However, whilst the theory of
dislocations and crystal defects has become a major stream in the
science of solid state, the corresponding effort applied to this
problem in amorphous materials must be considered rather small by
comparison.
\newline
The molecular theory of deformation kinetics came from a different
stream of science than that of structure and motion of crystal
defects (in particular dislocations). Its roots stretch to the
developmental stages of theories of chemical reactions and
thermodynamic description of their temperature dependence,
culminating in the key formulation by Arrhenius of the equation
for reaction rates. By the beginning of this century the concept
of activation entropy was included in the model, and it was
considered that molecules go both in the forward direction
(product state) and in the backward direction (reactant state).
\newline The development of statistical mechanics, and later quantum
mechanics, led to the concept of the potential energy surface.
This was a very important step in our modem understanding of
atomic models of deformation. Eyring's contribution to this
subject was the formal development of the transition state theory
which provided the basis for deformation kinetics, as well as all
other thermally activated processes, such as crystallisation,
diffusion, polymerisation. etc. [10] \newline
The motion of atoms is represented in the configuration space; on
the potential surface the stable molecules are in the valleys,
which are connected by a pass that leads through the saddle point.
An atom at the saddle point is in the transition (activated)
state. Under the action of an applied stress the forward velocity
of a (plastic) flow unit is the net number of times it moves
forward, multiplied by the distance it jumps. Eyring proposed a
specific molecular model of the amorphous structure and a
mechanism of deformation kinetics [10]. With reference to this
idea, this mechanism results in a (shear) strain rate given by
\begin{equation}  
 \dot{\Gamma}=2\frac{V_h}{V_m}\frac{k_B T}{h}\exp (\frac{-\Delta
E}{k_B T}) \sinh(\frac{V_h \tau}{2 k_B T})
\end{equation}
where $$V_h=\lambda_2\lambda_3\lambda, \hspace*{24mm}
V_m=\lambda_2\lambda_3\lambda_1,$$ $\lambda_1$ is the
perpendicular distance between two neighboring layers of molecules
sliding past each other, $\lambda$ is the average distance between
equilibrium positions in the direction of motion, $\lambda_2$ is
the distance between neighboring molecules in this same direction
(which may or may not equal $\lambda$), $\lambda_3$ is the
molecule to molecule distance in the plane normal to the direction
of motion, and $\tau$ is the local applied stress, $\Delta E$ is
the activation energy, $h$ is the Planck constant, $k_B$ is the
Boltzmann constant, $T$ is the temperature, $V_h$ is the
activation volume for the molecular event [9-10]. The deformation
kinetics of the chain is envisaged as the propagation of kinks in
the molecules into available holes. In order for the motion of the
kink to result in a plastic flow, it must be raised (energised)
into the activated state and pass over the saddle point. This was
the earliest molecular theory of yield behaviour in amorphous
materials, and Eyring presented a theoretical framework which
formed the basis of many subsequent considerations.
\newline
Solving Eqn. (1) for the force or $\tau$, one obtains:
\begin{equation}
  \tau=\frac{2 k_B T}{V_h} \sinh^{-1} (\frac{\dot{\Gamma}}{B}),
\end{equation}
which in the limit of small $(\dot{\Gamma}/B)$ reduces to Newton's
law for viscous deformation kinetics.
\newline
\noindent We  consider a steady deformation kinetics of the glassy
material in a wavy-rough microtube of $r_o$ (in mean-averaged
outer radius) with the outer wall being a fixed
wavy-rough surface : $r=r_o+\epsilon \sin(k \theta)$ 
where $\epsilon$ is the amplitude of the (wavy) roughness, and the
wave number : $k=2\pi /L $ ($L$ is the wave length). The schematic
is illustrated in Fig. 2. Firstly, this material can be expressed
as [9-10,14]
 $\dot{\Gamma}=\dot{\Gamma}_0  \sinh(\tau/\tau_0)$,
where $\dot{\Gamma}$ is the shear rate, $\tau$ is the shear
stress, and
\begin{equation}
\dot{\Gamma}_0 \equiv B= \frac{2 k_B T}{h}\frac{V_h}{V_m} \exp
(\frac{-\Delta E}{k_B T}),
\end{equation}
 is a function of temperature with the
dimension of the shear rate,
\begin{equation}
\tau_0 =\frac{2 k_B T}{V_h}
\end{equation}
 is the referenced (shear) stress, (for
small shear stress $\tau \ll \tau_0$, the linear dashpot
constitutive relation  is recovered and $\tau_0/\dot{\Gamma}_0$
represents the viscosity of the material). In fact, the force
balance gives the shear stress at a radius $r$ as $\tau=-(r
\,dp/dz)/2$. $dp/dz$ is the pressure gradient along the tube-axis
or $z$-axis direction.\newline Introducing the forcing parameter
$\Pi = -(r_o/2\tau_0) dp/dz$
then we have
 $\dot{\Gamma}= \dot{\Gamma}_0  \sinh ({\Pi r}/{r_o})$.
As the (shear) strain rate is
\begin{equation}
\dot{\Gamma}= \frac{du}{dr}
\end{equation}
($u$ is the rate of deformation (or velocity)  in the longitudinal
($z$-)direction of the microtube), after integration, we obtain
\begin{equation}
 u=u_s +\frac{\dot{\Gamma}_0 r_o}{\Pi} [\cosh \Pi - \cosh (\frac{\Pi r}{r_o})],
\end{equation}
here, $u_s$ is the rate of deformation or velocity over the
surfaces of the microtube, which is determined by the boundary
condition. We noticed that
 a general boundary condition for material deformation kinetics over a solid
surface was proposed (cf., e.g.,  [14]) as
\begin{equation}
 \delta u=L_s^0 \dot{\Gamma}
 (1-\frac{\dot{\Gamma}}{\dot{\Gamma}_c})^{-1/2},
\end{equation}
where $\delta u$ is the rate of deformation (or velocity) jump
over the solid surface, $L_s^0$ is a constant slip length and
$\dot{\Gamma}_c$ is the critical shear rate at which the slip
length diverges. The value of $\dot{\Gamma}_c$ is a function of
the corrugation of interfacial energy. We remind the readers that
this expression is based on the assumption of the shear rate over
the solid surface being much smaller than the critical shear rate
of $\dot{\Gamma}_c$. $\dot{\Gamma}_c$ represents the maximum shear
rate the material can sustain beyond which there is no additional
momentum transfer between the wall and material-flow molecules.
How generic this behavior is and whether there exists a comparable
scaling for glassy or amorphous materials remain open
questions.\newline At small pressure gradient, the shear-thinning
matter behaves like a Newtonian flow, while at high pressure
gradient, the shear-thinning matter flows in a plug-flow type.
Such a behavior is due to the shear thinning of the material,
i.e., the higher the shear rate is, the smaller is the (plastic)
flow resistance [7]. On the microscale, this shear-thinning matter
can bridge the Newtonian deformation kinetics to that of the
pluglike type and offers us a mechanistic model to study the
deformation kinetics in micro- and even nanodomains
 using the technique of continuum
mechanics.
\newline 
With the boundary condition from (cf., e.g.,  [18]), we shall
derive the rate of deformation (or velocity) field or deformation
kinetics along the wavy-rough microtube below using the boundary
perturbation technique (cf. [23]) and dimensionless analysis. We
firstly select the hydrodynamical diameter $L_r$ to be the
characteristic length scale and set
\begin{equation}
r'=r/L_r, \hspace*{6mm} R_o=r_o/L_r, 
\hspace*{6mm} \epsilon'=\epsilon/L_r.
\end{equation}
After this, for simplicity, we drop all the primes. It means, now,
$r$, $R_o$, $R_i$, and $\epsilon$ become dimensionless. The wall
is prescribed as $r=R_o+\epsilon \sin(k\theta)$, 
and the presumed fully-developed plastic flow is along the
$z$-direction (microtube-axis direction). Along the confined
(wavy) boundaries, we have the strain rate
\begin{equation}
 \dot{\Gamma}=(\frac{d u}{d n})|_{{\mbox{\small on surface}}},
\end{equation}
where, $n$ means the  normal. Let the rate of deformation $u$ be
expanded in $\epsilon$ :
\begin{equation}
 u= u_0 +\epsilon u_1 + \epsilon^2 u_2 + \cdots,
\end{equation}
and on the boundary, we expand $u(r_0+\epsilon dr,
\theta(=\theta_0))$ into
\begin{displaymath}
u(r,\theta) |_{(r_0+\epsilon dr ,\theta_0)}
=u(r_0,\theta)+\epsilon [dr \,u_r (r_0,\theta)]+ \epsilon^2
[\frac{dr^2}{2} u_{rr}(r_0,\theta)]+\cdots=
\end{displaymath}
\begin{equation}
 \hspace*{12mm} \{u_{slip} +\frac{\dot{\Gamma} R_o}{\Pi} \cosh
 (\frac{\Pi \bar{r}}{R_o})|_r^{R_o+\epsilon \sin(k\theta)},
 \hspace*{6mm} r_0 \equiv R_o;
\end{equation}
where the subscript means the partial differentiation (say, $u_r
\equiv
\partial u/\partial r$) and
\begin{equation}
 u_{slip}|_{{\mbox{\small on surface}}}=L_s^0 \dot{\Gamma} [(1-\frac{\dot{\Gamma}}{\dot{\Gamma}_c})^{-1/2}]
 |_{{\mbox{\small on surface}}},
\end{equation}
\begin{equation}
 u_{{slip}_0}= L_s^0 \dot{\Gamma}_0 [\sinh\Pi(1-\frac{\dot{\Gamma}_0 \sinh\Pi}{
 \dot{\Gamma}_c})^{-1/2}].
\end{equation}
Now, on the outer wall (cf., e.g., [23]), the (shear) strain rate 
\begin{displaymath}
 \dot{\Gamma}=\frac{du}{dn}=\nabla u \cdot \frac{\nabla (r-R_o-\epsilon
\sin(k\theta))}{| \nabla (r-R_o-\epsilon \sin(k\theta))
|}=[1+\epsilon^2 \frac{k^2}{r^2}  \cos^2 (k\theta)]^{-\frac{1}{2}}
[u_r |_{(R_o+\epsilon dr,\theta)} -
\end{displaymath}
\begin{displaymath}  
 \hspace*{12mm} \epsilon \frac{k}{r^2}
\cos(k\theta) u_{\theta} |_{(R_o+\epsilon dr,\theta)}
]=u_{0_r}|_{R_o} +\epsilon [u_{1_r}|_{R_o} +u_{0_{rr}}|_{R_o}
\sin(k\theta)-
\end{displaymath}
\begin{displaymath}
  \hspace*{12mm}  \frac{k}{r^2} u_{0_{\theta}}|_{R_o} \cos(k\theta)]+\epsilon^2 [-\frac{1}{2} \frac{k^2}{r^2} \cos^2
(k\theta) u_{0_r}|_{R_o} + u_{2_r}|_{R_o} + u_{1_{rr}}|_{R_o} \sin(k\theta)+ 
\end{displaymath}
\begin{equation}
   \hspace*{12mm} \frac{1}{2} u_{0_{rrr}}|_{R_o} \sin^2 (k\theta) -\frac{k}{r^2}
\cos(k\theta) (u_{1_{\theta}}|_{R_o} + u_{0_{\theta r}}|_{R_o}
\sin(k\theta) )] + O(\epsilon^3 ) .
\end{equation}
Considering $L_s^0 \sim R_o \gg \epsilon$ case, we presume
$\sinh\Pi \ll \dot{\Gamma}_c/\dot{\Gamma_0}$ so that we can
approximately replace
$[1-(\dot{\Gamma}_0 \sinh\Pi)/\dot{\Gamma}_c]^{-1/2}$
by
$[1+\dot{\Gamma}_0 \sinh\Pi/(2 \dot{\Gamma}_c)]$.
With equations (6),(7),(9), (10), (11) and (14), using the
definition of the (shear) strain rate $\dot{\Gamma}$, we can
derive the rate of deformation (or velocity) field up to the
second order. The key point is to firstly obtain the slip rate of
deformation (or velocity) along the wavy boundaries or surfaces.
\newline After lengthy mathematical manipulations and using
 $(1-{\dot{\Gamma}}/{\dot{\Gamma}_c})^{-1/2}\approx 1+{\dot{\Gamma}}/({2
 \dot{\Gamma}_c})$,
\begin{equation}
 u_0=-\frac{\dot{\Gamma}_0 R_o}{\Pi} [\cosh
 (\frac{\Pi r}{R_o})-\cosh \Pi]+u_{{slip}_0},
\end{equation}
\begin{equation}
  u_1=
 \dot{\Gamma}_0 \sin (k\theta) \sinh \Pi +u_{{slip}_1},
\end{equation}
we have %
\begin{displaymath}
 u_{slip}=L_s^0 \{[-u_{0_r}(1-\frac{u_{0_r}}{2
 \dot{\Gamma}_c})]|_{r=R_o}+\epsilon
 [-u_f(1-\frac{u_{0_r}}{\dot{\Gamma}_c})]|_{r=R_o}+\epsilon^2
 [\frac{u_f^2}{2\dot{\Gamma}_c}-u_{sc}
 (1-\frac{u_{0_r}}{\dot{\Gamma}_c})]|_{r=R_o}\}=
\end{displaymath}
\begin{equation}
 \hspace*{36mm} u_{slip_0} +\epsilon \,u_{slip_1} + \epsilon^2 u_{slip_2}
 +O(\epsilon^3)
\end{equation}
where
\begin{equation}
 u_{0_r}= -\dot{\Gamma}_0 \sinh(\frac{\Pi}{R_o}r),
\end{equation}
\begin{equation}
 u_{0_{rr}}=-\dot{\Gamma}_0 \frac{\Pi}{R_o} \cosh(\frac{\Pi}{R_o}r),
\end{equation}
\begin{equation}
 u_{0_{rrr}}=-\dot{\Gamma}_0
\frac{\Pi^2}{R_o^2}\sinh(\frac{\Pi}{R_o}r),
\end{equation}
\begin{equation}
 u_f =u_{1_r} + u_{0_{rr}} \sin (k\theta)-\frac{k}{r^2} \cos (k\theta)
 u_{0_{\theta}}=-\frac{\Pi}{R_o}\dot{\Gamma}_0 \cosh(\frac{\Pi}{R_o}r) \, \sin (k\theta),
\end{equation}
and
\begin{equation}
 u_{sc} =-\frac{k^2}{2 r^2}\cos^2 (k \theta) u_{0_r} +\frac{1}{2}
 u_{0_{rrr}} \sin^2 (k\theta)=\frac{1}{2}\dot{\Gamma}_0 [\frac{k^2}{2 r^2}\cos^2 (k \theta)
 -\frac{\Pi^2}{R_o^2}\sin^2 (k\theta)]\sinh(\frac{\Pi}{R_o}r).
\end{equation}
Thus, at $r=R_o$, up to the second order,
\begin{displaymath}
 u_{slip}\equiv u_s=L_s^0 \dot{\Gamma}_0  \sinh\Pi(1+\frac{K_0}{2})+\epsilon \dot{\Gamma}_0 \sin(k\theta)
 [\sinh \Pi+ \frac{\Pi}{R_o}L_s^0\cosh\Pi \, (1+K_0)]+\epsilon^2 L_s^0\frac{\dot{\Gamma}_0 }{2}\{
 [
\end{displaymath}
\begin{equation}
 \frac{\Pi \cosh \Pi}{R_o L_s^0}  \sin^2 (k\theta)-\frac{k^2}{R_o^2} \cos^2 (k\theta)+
 \frac{\Pi^2}{R_o^2} \sin^2 (k\theta)]\sinh\Pi (1+K_0) +
 \frac{\Pi^2}{R_o^2} \frac{\dot{\Gamma}_0}{\dot{\Gamma}_c}
  \cosh^2 \Pi \,\sin^2 (k\theta) \},
\end{equation}
where
\begin{equation}
K_0=1+({\dot{\Gamma}_0
 \sinh\Pi})/{\dot{\Gamma}_c}
\end{equation}
From the rate of deformation (or velocity) fields (up to the
second order), we can integrate them with respect to the
cross-section to get the volume (plastic) flow rate ($Q$, also up
to the second order here).
\begin{equation}
  Q=\int_0^{\theta_p} \int^{R_o+\epsilon \sin(k\theta)}
 u(r,\theta) r
 dr d\theta =Q_{smooth} +\epsilon\,Q_{p_0}+\epsilon^2\,Q_{p_2}.
\end{equation}
\section{Results and discussion}
We firstly check the roughness effect upon the shearing
characteristics because there are no available experimental data
and numerical simulations for the same geometric configuration
(microscopic tubes with wavy corrugations in transverse
direction). With a series of forcings (due to imposed pressure
gradients) : $\Pi\equiv R_o (-dp/dz)/(2\tau_0)$, we can determine
the enhanced shear rates ($d\Gamma/dt$) due to forcings. From
equation (5), we have (up to the first order)
\begin{equation}
 \frac{d\Gamma}{dt}=\frac{d\Gamma_0}{dt} [ \sinh \Pi+\epsilon
 \sin(k\theta) \frac{\Pi}{R_o} \cosh \Pi].
\end{equation}
The calculated results are demonstrated in Figs. 3 and 4. The
parameters are fixed below (the orientation effect :
$\sin(k\theta)$ is fixed here). $r_o$ (the mean outer radius) is
selected as the same as the slip length $L_s^0=100$ nm. The
amplitude of wavy roughness is $\epsilon=0.04, 0.07, 0.1$,  the
Boltzmann constant ($k_B$) is $1.38 \times 10^{-23}$
Joule/$^{\circ}$K, and the Planck constant ($h$) is $6.626 \times
10^{-34}$ Joule $\cdot$ s. \newline \noindent In each panel, the
inner curve is the relevant boundary of the tube or the geometric
part of the presentation. The distance between the inner and
corresponding outer curves is the calculated physical shear rate :
$\dot{\Gamma}$.
We can observe once the temperature ($T$) changes a little  from
0.1 $^{\circ}$ to 0.15 $^{\circ}$, the enhancement of
$\dot{\Gamma}$ becomes at least three orders of magnitude (for
$\Pi=1$, the activation energy : $3\times 10^{-23}$ Joules). Even
at very low temperature Fig. 4 gives very large strain rates which
are required to to obtain the
 necessary strain for plastic deformation.
 Thus, the constitutive relations is highly nonlinear
 at rather low temperature regime [9].
 It is worth pointing out that the Eyring model requires the
interaction between atoms in the direction perpendicular to the
shearing direction for the momentum transfer. This might explain
why our result is orientation dependent. The effect of
wavy-roughness will be significant once the forcing ($\Pi$) is
rather large (the maximum is of the order of magnitude of
$\epsilon [\Pi \tanh(\Pi)/R_o]$).
\newline
\noindent To be specific, we can illustrate the shear rate
($\dot{\Gamma}$) with respect to the temperature ($T$) once we
calculate $\dot{\Gamma}_0$ as the latter is temperature dependent
(but presumed roughness independent here) which could be traced
from equation (1). This is shown in Fig. 5. \newline Note that,
based on the rate-state Eyring model (of stress-biased thermal
activation), structural rearrangement is associated with a single
energy barrier (height) $E$ that is lowered or raised linearly by
a (shear) yield stress $\tau$. If the transition rate is
proportional to the plastic (shear) strain rate (with a constant
ratio : $C_0$; $\dot{\Gamma} =C_0 R_t$, $R_t$ is the transition
rate in the direction aided by stress), we have $\tau = E / V^*+(
k_B T / V^*)
 \ln ( \dot{\Gamma} /C_0 \nu_0)$ or
\begin{equation}
\tau = \frac{E}{V^*}+(\frac{ k_B T}{V^*})
 \ln (\frac{|\dot{\tau}| V^*}{\nu_0 k_B T}),
\end{equation}
where $V^* \equiv V_h$ is a constant called the activation volume,
$k_B$ is the Boltzmann constant, $T$ is the temperature, $\nu_0$
is an attempt frequency or transition rate [8,25], and
$\dot{\tau}$ is the stress rate. Normally, the value of $V^*$ is
associated with a typical volume required for a molecular shear
rearrangement. Thus, if there is a rather-small (plastic) flow (of
the glass) at low temperature environment then it could be related
to a barrier-overcoming or tunneling for shear-thinning matter
along the wavy-roughness (geometric valley and peak served as
atomic potential surfaces) in cylindrical micropores when the
wavy-roughness is present. Once the geometry-tuned potentials
(energy) overcome this barrier, then the tunneling (spontaneous
transport) inside wavy-rough cylindrical micropores occurs.
\newline
\noindent To examine the behavior of the shear rate at low
temperature regime, we calculate $\dot{\Gamma}_0$ and
$\dot{\Gamma}$ ($C_0 \nu_0=5\times 10^{10}$ s$^{-1}$) with respect
to the temperature $T$ and show the results in Fig. 5. For a
selected activation energy : $5\times 10^{-24}$ Joule or $\sim
10^{-5}$ eV (a little bit smaller than the binding energy of
$^3$He), we can find a sharp decrease of shear rates around $T\sim
0.01 ^{\circ}$K. Below this temperature, there might be nearly
frictionless transport of glassy matter. Note also that, according
to Cagle and Eyring [8], $V^*=3 V \delta \Gamma/2$ for certain
material during an activation event, where $V$ is the deformation
volume, $\delta\Gamma$ is the increment of shear strain.\newline
If we select a (fixed) temperature, say, $T=0.1 ^{\circ}$K, then
from the expression of $\tau_0$, we can obtain the shear stress
$\tau$ corresponding to above forcings ($\Pi$) :
\begin{equation}
 \tau =\tau_0 \sinh^{-1} [\sinh(\Pi)+\epsilon
 \sin(k\theta) \frac{\Pi}{R_o} \cosh(\Pi)].
\end{equation}
There is no doubt that the orientation effect ($\theta$) is also
present for deformation kinetics of amorphous matter. For
illustration (shown in Fig. 6), we only consider the maximum case
: $|\sin(k\theta)|=1$. The trend of enhancement due to $\Pi$
(pressure-forcing) and $\epsilon$ (roughness) is similar to those
presented in Figs. 3 and 4. We remind the readers that, due to the
appearance of $\tau_0$, we fix the temperature to be $0.1
^{\circ}$K and the activation volume : $10^{-25}$ m$^3$.\newline
\noindent In fact, as shown in Fig. 6, the calculated (shear)
stress (which is directly linked to the resistance of the glassy
matter) also shows a sudden decrease around $T\sim 0.05 ^{\circ}$K
especially for the case of ($C_0 \nu_0=2\times 10^{9}$ s$^{-1}$).
Here, the activation volume ($V^*$ or $V_h$) is selected as $0.2$
nm$^3$ [19]. Thus, the nearly frictionless transport of the glassy
fluid at low temperature environment (relevant to the
supersolidity, cf. [20]) could be related to a barrier-overcoming
or tunneling for shear-thinning matter along the wavy-roughness
(geometric valley and peak served as atomic potential surfaces) in
cylindrical micropores when the wavy-roughness is present. Once
the geometry-tuned potentials (energy) overcome this barrier, then
the tunneling (almost frictionless transport) inside
wavy-rough cylindrical micropores occurs. \newline 
\noindent We also noticed that, as described in [9-10], mechanical
loading lowers energy barriers, thus facilitating progress over
the barrier by random thermal fluctuations. The simplified Eyring
model approximates the loading dependence of the barrier height as
linear. This Eyring model, with this linear barrier height
dependence on load, has been used over a large fraction of the
last century to describe the response of a wide range of systems
and underlies modern approaches to sheared glasses. The linear
dependence will always correctly describe small changes in the
barrier height, since it is simply the first term in the Taylor
expansion of the barrier height as a function of load. It is thus
appropriate when the barrier height changes only slightly before
the system escapes the local energy minimum. This situation occurs
at higher temperatures; for example, Newtonian deformation
kinetics is obtained in the Eyring model in the limit where the
system experiences only small changes in the barrier height before
thermally escaping the energy minimum. As the temperature
decreases, larger changes in the barrier height occur before the
system escapes the energy minimum (giving rise to, for example,
non-Newtonian deformation kinetics). In this regime, the linear
dependence is not necessarily appropriate, and can lead to
inaccurate modelling. This explains why we should adopt the
hyperbolic sine law [9-10] to treat the glassy matter.
\newline
To be specific, our results are rather sensitive to the
temperature ($T$) and the activation energy. Fig. 7 shows
especially the temperature dependence of the forcing parameter
($\Pi$) if $dp/dz$ is prescribed (say, around $6\times 10^{10}$
Pa/m) and the activation volume is $0.2$ nm$^3$ ($r_o=100$ nm). We
can observe that once $T$ increases $\Pi$ decreases.
$\dot{\Gamma}$ calculated using prescribed $\Pi$ and using
directly $T$ also differs. \newline Finally, we present the
calculated maximum velocity (unit : m/s) with respect to the
temperature in Fig. 8. Geometric parameters : $r_o$ and the
activation volume are the same as those in Fig. 7 and the
roughness amplitude $\epsilon=0.02, 0.05 R_o$. We consider the
effect of the activation energy : $1.5 \times 10^{22}$ and
$2\times 10^{22}$ Joule. Around $T\sim 0.35 ^{\circ}$K, the
maximum velocity (of the glassy matter) either keeps decreasing as
the temperature increases for larger activation energy  or instead
increases as the temperature increases for smaller activation
energy! The results presented in Fig. 8 might be related to the
microscopic origin for physical aging  or effects of thermal
history [21] and indeterminate solutions discussed in [22]. The
latter observation might be related to the argues raised in [23]
for the annealing process of solid helium at similar low
temperature environment if we treat the solid helium to be glassy
at low temperature regime. On the other hand, we like to remind
the readers about the role of roughness again. If $^4$He is
isotropic (prepared), and the impurity of $^3$He distribution was
added or mixed into $^4$He quite regularly (equally-distributed).
It's possible that the role of $^3$He [24] plays under certain
selected activation energy and activation volume is qualitatively
the same as that of the roughness here. In general, as commented
in [25] : {\it An understanding of the NCRI effect may therefore
require explicit treatment of the effects of strain on the crystal
properties $\cdots$ We have used a narrow x-ray beam to study the
defects in $^4$He crystals, and we find crystals contain a
microstructure of mosaic regions consistent with small angle grain
boundaries $\cdots$}, our presentation here is crucial to the
understanding of strain upon the  solid helium at rather low
temperature regime once the solid helium is presumed to be glassy
(cf. [26] for the detailed references) within confined
microdomains. To give our approach another comparison with the
transport data in [3], e.g., with the same temperature $0.4$ K as
in Fig. 4 of [3], once we set the activation energy ($\Delta E$)
to be 1.0 $\times 10^{-22}$ Joule (or $\sim 10^{-3}$ eV; cf. Fig.
3 in [27]), the radius to be $\sim 1 \mu$m, with other parameters
the same as those in Fig. 8 above, then we have the maximum
velocity being around $100 \mu$m/sec which is almost the same as
that estimated in [3] (where the uniform flow was presumed and the
fraction of the helium that can flow is $2.5 \times 10^{-6}$).
This is illustrated in Fig. 9.
\section{Conclusions}
To conclude in brief, we analytically obtain a class of
temperature as well as activation energy dependent fields of the
rate of deformation for glassy material (like solid helium) in
microscopically confined wavy-rough domain at very low temperature
regime. The effects of wavy corrugation upon the confined
deformation kinetics at very-low temperature are clearly
illustrated. It is found that there exist almost frictionless
plastic flow fields for the rate of deformation of glassy material
inside cylindrical micropores at very low temperature once the
micropore surface is wavy-rough and the activation energy is
prescribed. Once the temperature, activation volume, and geometry
are fixed, the increase of activation energy instead reduces
significantly the (maximum) rate of deformation of the glassy
matter. The critical rate of deformation is proportional to the
(referenced) shear rate, the slip length, the orientation and the
amplitude of the wavy-roughness as illustrated above. Our
approaches can recover those transport results reported in [3] for
the same physical parameters.
\newline
\subsection*{Acknowledgement.} The author is partially supported
by the 2007-Hebei-NU Starting Funds for Scientific Researcher. The
author thanks the hospitality for the Visiting-Scholar Program of
the Chern Shiing-Shen Institute of Mathematics, Nankai University.
  \newline
%
%


\newpage

\psfig{file=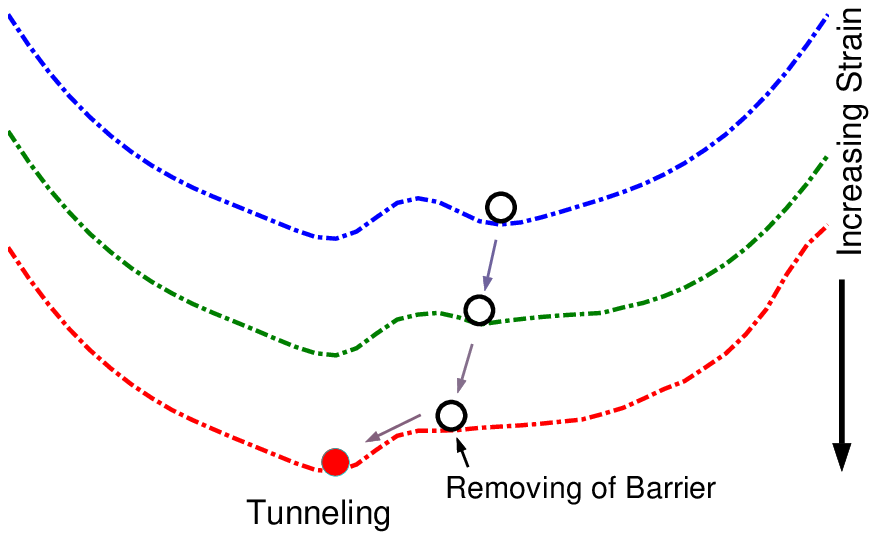,bbllx=-1.0cm,bblly=19cm,bburx=10cm,bbury=26cm,rheight=7cm,rwidth=7cm,clip=}

\begin{figure}[h]
\hspace*{10mm} Fig. 1 Increasing strain causes a local energy
minimum to flatten until it disappears \newline \hspace*{10mm}
(removing of energy barrier or quantum-like tunneling). The
structural contribution to \newline \hspace*{10mm} the shear
stress is shear thinning.
\end{figure}

\newpage

\psfig{file=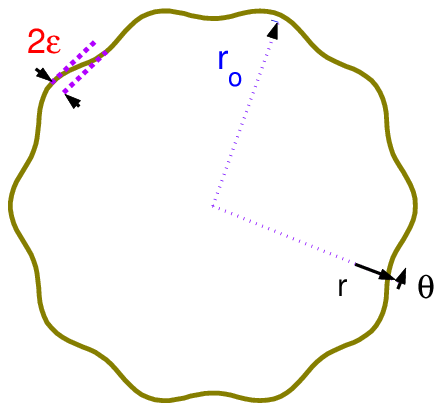,bbllx=1.0cm,bblly=16.5cm,bburx=18cm,bbury=25cm,rheight=6cm,rwidth=6cm,clip=}

\begin{figure}[h]
\hspace*{6mm} Fig. 2.  Schematic  of a cylindrical micropore.
 $\epsilon$ is
the amplitude of small wavy-roughness.
\end{figure}

\newpage

\psfig{file=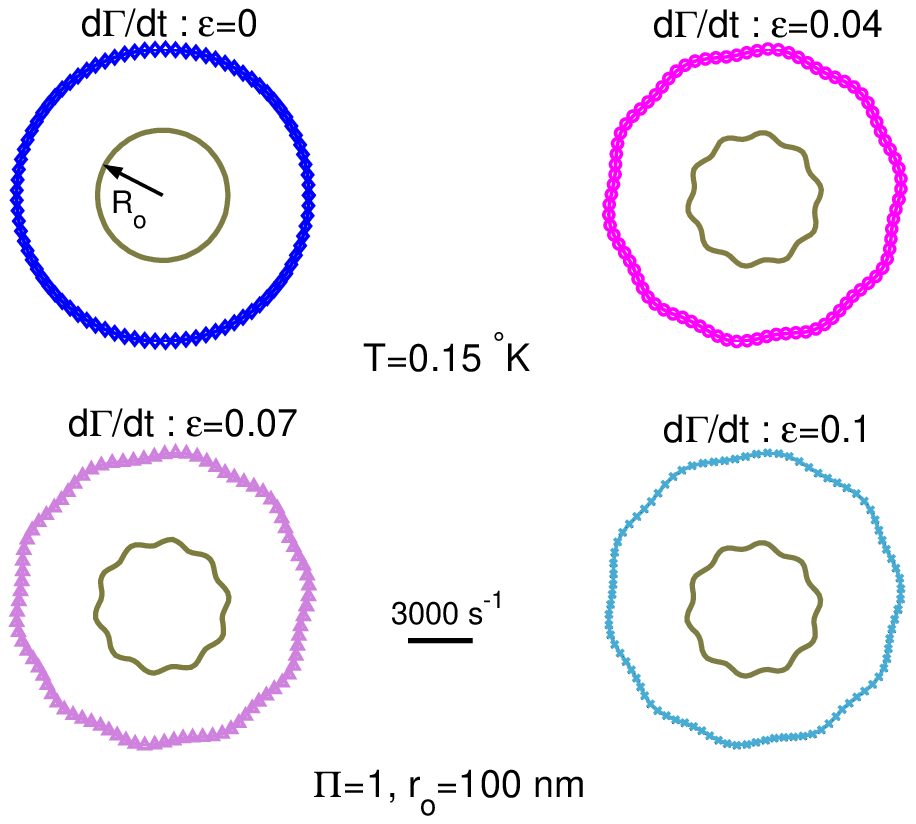,bbllx=-1.0cm,bblly=14.2cm,bburx=16cm,bbury=26cm,rheight=10cm,rwidth=7.6cm,clip=}

\begin{figure}[h]
{\small Fig. 3. Comparison of the shear rate ($\dot{\Gamma}$) of
glassy matter in smooth microtubes and wavy-rough microtubes for
$k=10$, $\epsilon=0.0, 0.04, 0.07, 0.1$ with $r_o$=100 nm.
$\dot{\Gamma_c}/\dot{\Gamma}_0=10$ and $L^0_s=r_o$. $k$ is the
wave number and $\epsilon$ is the amplitude of the wavy-roughness.
$T$ is the temperature. The solid-line length represents the scale
of $\dot{\Gamma}=10 s^{-1}$.}

\end{figure}

\newpage

\psfig{file=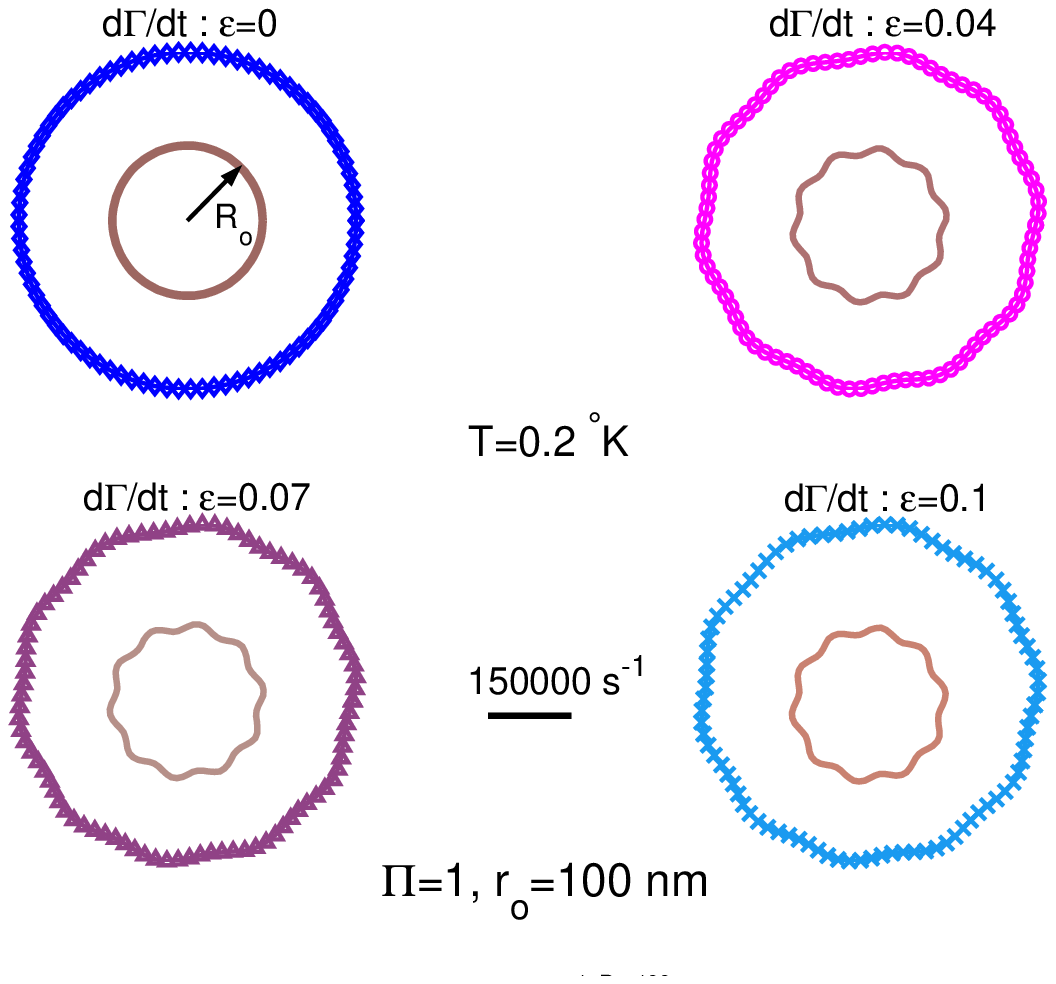,bbllx=-1.0cm,bblly=14cm,bburx=14cm,bbury=26.4cm,rheight=10.6cm,rwidth=10cm,clip=}
\begin{figure}[h]
{\small Fig. 4. Comparison of the shear rate ($\dot{\Gamma}$) of
glassy matter in smooth microtubes and wavy-rough microtubes for
$k=10$, $\epsilon=0.0, 0.04, 0.07, 0.1$ with $r_o$=100 nm.
$\dot{\Gamma_c}/\dot{\Gamma}_0=10$ and $L^0_s=r_o$. $k$ is the
wave number and $\epsilon$ is the amplitude of the wavy-roughness.
$T$ is the temperature. The solid-line length represents the scale
of $\dot{\Gamma}=3000 s^{-1}$.} As the temperature increases a
little, $\dot{\Gamma}$ increases significantly.

\end{figure}

\newpage

\psfig{file=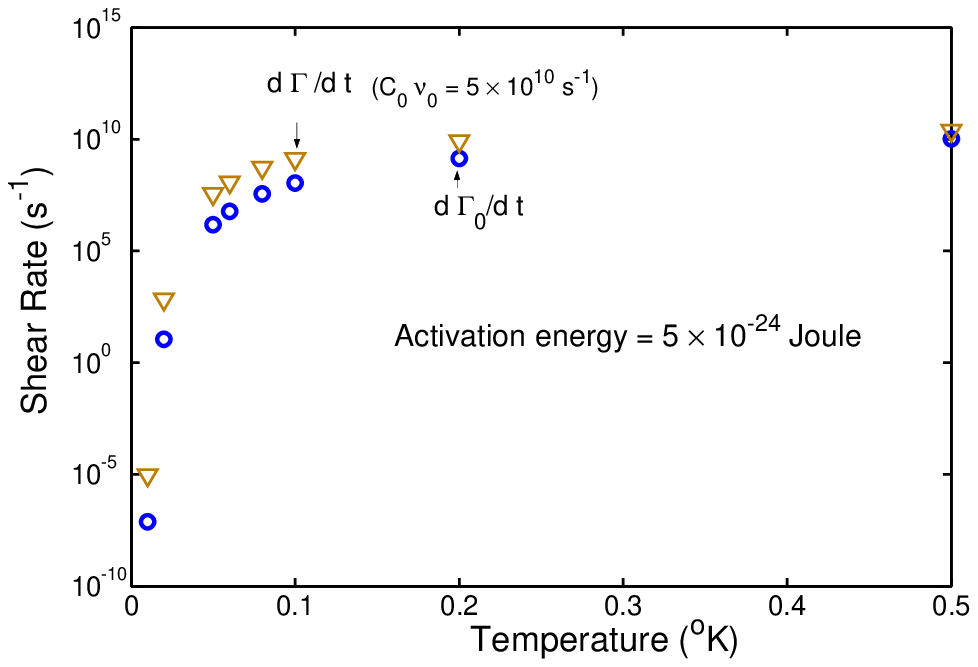,bbllx=-1.0cm,bblly=18cm,bburx=12cm,bbury=25.6cm,rheight=7.6cm,rwidth=9cm,clip=}

\begin{figure}[h]
\hspace*{10mm} Fig. 5 Comparison of calculated shear (strain)
rates using an activation energy $5\times 10^{-24}$ \newline
\hspace*{10mm} Joule or $\sim 10^{-5}$ eV. There is a sharp
decrease of shear rate around $T\sim 0.01 ^{\circ}$K.
\end{figure}

\newpage

\psfig{file=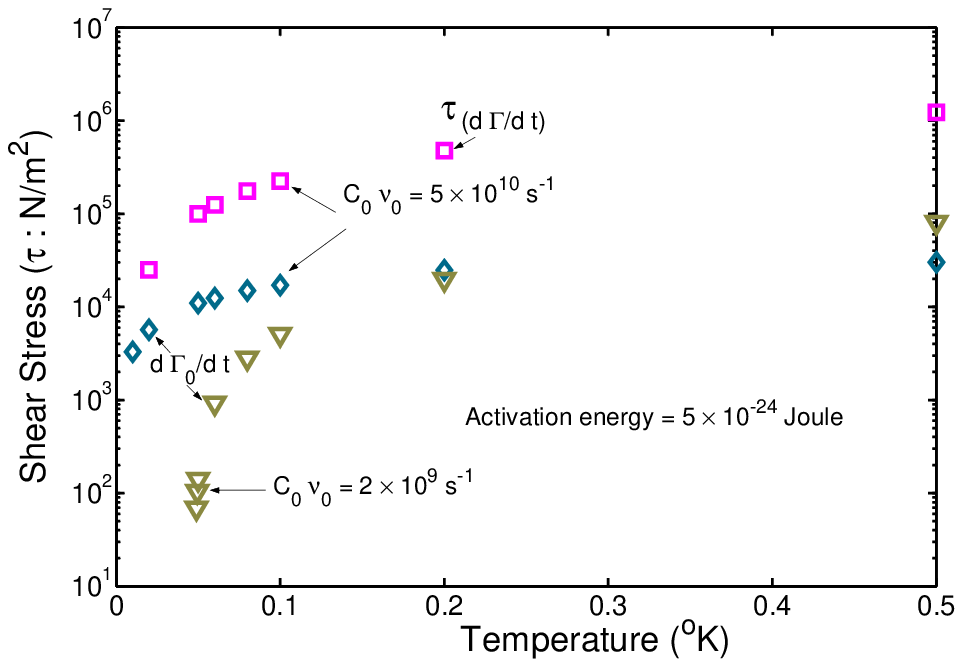,bbllx=-1.0cm,bblly=20.5cm,bburx=12cm,bbury=28.4cm,rheight=8cm,rwidth=8.5cm,clip=}

\begin{figure}[h]
\hspace*{10mm} Fig. 6  Comparison of calculated (shear) stresses
using an activation energy $5\times 10^{-24}$
\newline \hspace*{10mm} Joule or $\sim 10^{-5}$ eV. There is a
sharp decrease of shear stress around $T\sim 0.05 ^{\circ}$K for
\newline \hspace*{10mm}  $C_0 \nu_0=2\times 10^9$ s$^{-1}$. Below
$0.05  ^{\circ}$K, the transport of glassy matter is nearly
frictionless. \newline \hspace*{10mm} $\nu_0$ is an attempt
frequency or transition rate [10,19].
\end{figure}

\newpage

\psfig{file=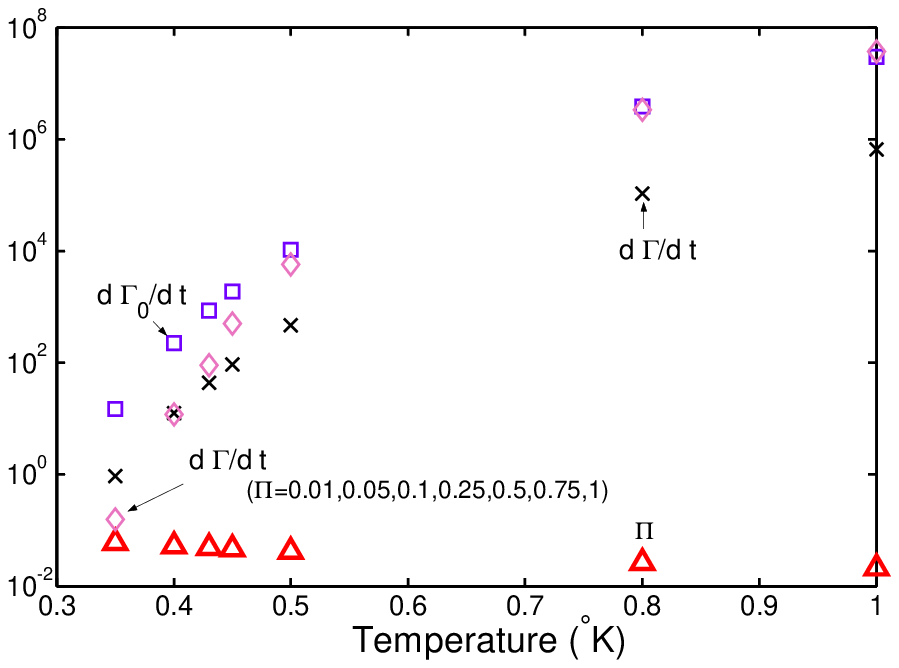,bbllx=-1.0cm,bblly=19cm,bburx=12cm,bbury=26.5cm,rheight=7.5cm,rwidth=9cm,clip=}

\begin{figure}[h]
\hspace*{10mm} Fig. 7 Comparison of calculated shear (strain)
rates using an activation energy $10^{-22}$ \newline
\hspace*{10mm} Joule or $\sim 10^{-3}$ eV.  The temperature ($T$)
dependence of $\Pi$ (forcing) is also demonstrated. \newline
\hspace*{10mm} Forcing ($\Pi$) decreases as the temperature ($T$)
increases. $\dot{\Gamma}$ (mark : diamond) calculated \newline
\hspace*{10mm} using prescribed $\Pi$
 is different from that (mark : cross)
using directly $T$ (temperature).
\end{figure}

\newpage

\psfig{file=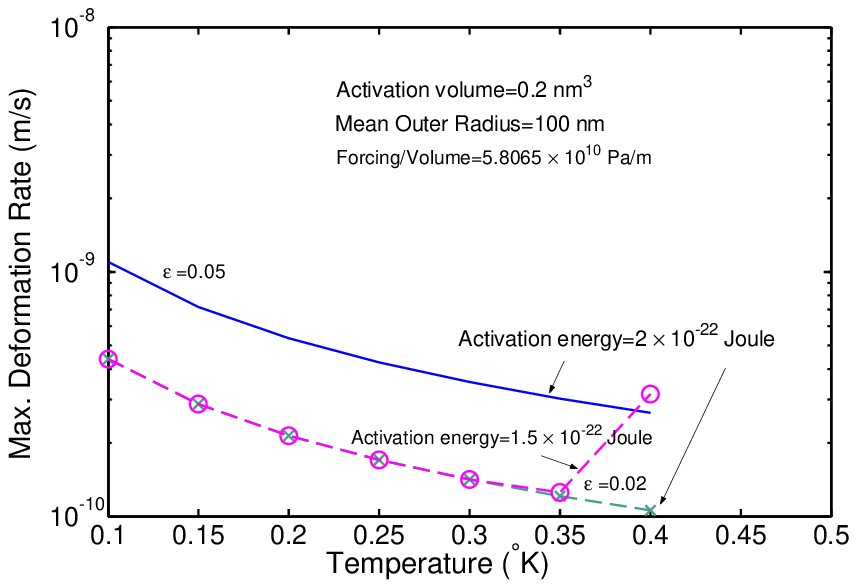,bbllx=-1.0cm,bblly=20.5cm,bburx=12cm,bbury=28.5cm,rheight=8cm,rwidth=9.5cm,clip=}

\begin{figure}[h]
\hspace*{10mm} Fig. 8  Comparison of calculated (maximum) velocity
(unit : m/s) using two activation \newline \hspace*{10mm} energies
$1.5 \times 10^{-22}$ and $2\times 10^{-22}$
 Joule. Around $T\sim 0.35 ^{\circ}$K, the monotonic trend of
  \newline \hspace*{10mm} velocity (or deformation rate)
   bifurcates as the temperature increases. \newline \hspace*{10mm}
  $r_o=100$ nm and $\epsilon=0.02, 0.05 R_o$.
\end{figure}

\newpage

\psfig{file=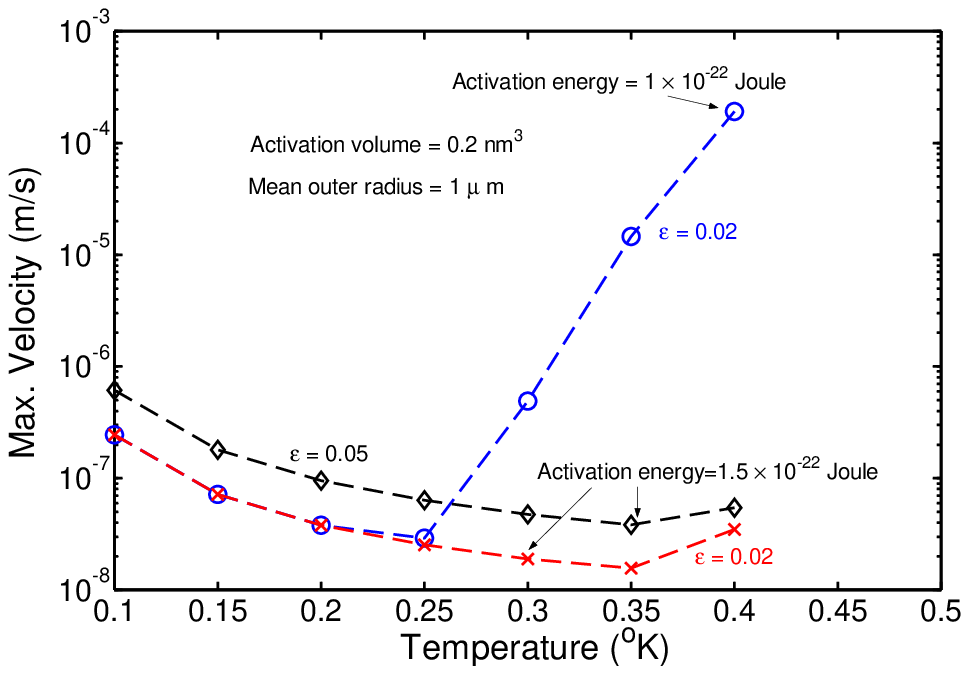,bbllx=-1.0cm,bblly=18cm,bburx=12cm,bbury=27cm,rheight=8.5cm,rwidth=10cm,clip=}

\begin{figure}[h]
\hspace*{10mm} Fig. 9  Comparison of calculated (maximum) velocity
(unit : m/s) using two activation \newline \hspace*{10mm} energies
$1 \times 10^{-22}$ and $1.5\times 10^{-22}$
 Joule. Around $T\sim 0.25$ or $0.35 ^{\circ}$K, the monotonic trend of
  \newline \hspace*{10mm} velocity (or deformation rate)
   bifurcates as the temperature increases. \newline \hspace*{10mm}
  $r_o=1 \mu$m and $\epsilon=0.02, 0.05 R_o$.
\end{figure}

\end{document}